\documentclass{aa}             
\usepackage{graphicx}
\usepackage{txfonts}
\begin{document}

\title{Hysteresis in spectral state transitions - a challenge for
theoretical modeling.}
\author{E. Meyer-Hofmeister, B.F. Liu and F. Meyer}
\offprints{Emmi Meyer-Hofmeister}
\institute{Max-Planck-Institut f\"ur Astrophysik, Karl-
Schwarzschildstr.~1, D-85740 Garching, Germany
} 

\date{Received: / Accepted:}

\abstract{Many low-mass X-ray binaries show both hard and soft spectral
states. For several sources the transitions between these states have been 
observed, mostly from the soft to the hard state during a luminosity
decrease. In a few cases also the transition from the hard to the soft 
state was observed, coincident with an increase of the luminosity. 
Surprisingly 
this luminosity was not the same as the one during a following change
back to the hard state. The values differed by a factor of about 3 to
5. We present a model for this hysteresis in the light curves of 
low-mass X-ray binaries (sources with neutron stars or black holes). We show
that the different amount of Compton cooling or heating acting on the accretion
disk corona at the time of the transition causes this switch in the 
accretion mode at different mass accretion rates and therefore
different luminosities. The inner disk during the soft state provides
a certain amount of Compton cooling which is either not present or much less 
if the inner region is filled with a hot advection-dominated
accretion flow (ADAF) that radiates a hard spectrum.

}
\titlerunning {}
\maketitle

\section{Introduction}

One of the most fascinating features found in X-ray binary
observations are the changes between a soft and a hard spectrum. 
Transitions between the two spectral states were observed for both neutron
star and black hole systems (Tanaka \& Shibazaki 1996). For the neutron 
star 1608-522 Mitsuda et al. (1989)  already observed the change from a soft 
to a hard state. One of the early observed spectral transitions in black hole
binaries was found for GRS 1124-684, Nova Mus, (Ebisawa et al. 1994)
Esin et al. (1997) modeled this spectral difference using the
concept of an inner advection-dominated accretion flow (ADAF).
The soft and hard spectral states are then understood as
originating from accretion via a disk which reaches inward to the
compact object in the soft state, or an advection-dominated hot
coronal flow/ADAF in the inner part and accretion via a disk only in the
outer regions in the hard state. 

This scheme of advection-dominated accretion introduced to 
model the very low luminosities observed (for a review and references
therein see Narayan et al. 1998) was further improved in
correspondence to the growing body of observations at different
wavelengths (Di Matteo  et al. 2000, for a review see Narayan 2002). 
The basic picture is clear, but the physics of the hot
coronal gas is complex and assumptions are unavoidable. Having
the two modes of accretion in mind an even more demanding question is what
determines the location in the disk where the mode of accretion changes
from disk accretion to the ADAF.

To study the change between the different accretion modes, low mass
X-ray binaries (LMXBs) that contain a neutron star or a black hole 
primary (reviews by Tanaka \& Shibazaki 1996, Chen et al. 1997, McClintock 
\& Remillard 2004) are suitable objects. The mass overflow rates from
the companion star are low,
and the disk becomes ionized and an outburst occurs only after mass
has accumulated for a long time.

Distinct from these sources are the persistently bright high 
mass X-ray binaries (HMXBs) with a high mass $\ge 10 M_\odot$
companion star, mostly an O or Be star (review by Charles \&
Coe 2004). These sources are wind accretors. Their luminosities vary
much less, but seems for at least the three sources, Cyg X-1, LMC X-1,
and LMC X-3 to be in the range which allows transitions.

There is a special feature in the appearance of the transition of 
spectral states: the hard-soft transition does not occur at the same luminosity
as the soft-hard transition. The latter seems to happen at a
luminosity lower by a factor of about 5 as far as can be seen from the 
observations. The aim of our paper is to give an explanation for this 
peculiar ``hysteresis''. 

In Sect.2 we put together the relevant observations.
Sect. 3 summarizes the theoretical work on spectral state
transitions. 
In Sect.4 we present our computational results.
The Compton effect that photons from the central light source have on
the coronal electrons depends on the spectrum of the radiation from
the innermost region and is different in the two cases, where either a disk 
exists when the transition from the soft to the hard spectrum occurs or a 
vertically extended hot flow when the transition from the hard to
the soft spectrum occurs.  We show that this  naturally leads to
different transition luminosities. In Sect.5 we show that this effect 
explains the hysteresis in the spectral transition 
luminosity. In Sect.6 we critically discuss previous suggestions  
for the cause of the hysteresis. Conclusions follow in Sect.7.

\section{Hysteresis in lightcurves of X-ray binaries - observations}
Spectral state transitions are observed
in three groups of sources: in HMXBs, in  black hole and in neutron star 
LMXBs. For Cyg X-1, the best known HMXB, mostly
in the hard state, a rich documentation is available. 
In this system no hysteresis is observed (see Sect. 6.2). For the other
persistent source, LMC X-3, mostly in the soft state, a  hard state
occurred in 1998 (Wilms et al. 2001). The X-ray novae have very rare
outbursts so that the chance to observe the hard to soft transition
during the rise to a new outburst is low.

Maccarone (2003) searched for observations of soft to hard
transitions. For his sample of 4 transient black hole X-ray sources, 
2 persistent black hole binaries and 3 neutron star binaries he finds 
transition luminosities in a narrow range of a few percent of the
Eddington luminosity. Kalemci et al. (2003) analyze
the PCA/RXTE data from all galactic black hole transients observed
with RXTE between 1996 and 2001 that made a state
transition during outburst decay. But the number of sources where we can
compare the luminosity for both transitions is small.

\subsection{Observations of individual sources}
A hysteresis in the spectral state transition was observed in the
following sources:

\it {Aql X-1:}\\
\rm{
For Aql X-1, the only neutron star LMXB considered here,
fortunately a full outburst cycle has been
observed. The luminosity at the 
hard-soft transition was found to be about 5 times higher than the
soft-hard transition luminosity (Maccarone \& Coppi 2003), 
4.2 to $5.5\times  10^{36}\rm{erg/s}$ in the rise 
of the outburst and 6.1-7.5 $\times 10^{35}\rm{erg/s}$ in the decline.
}

\it{GX 339-4:}\\		
\rm{
This was the system (together with GS 1124-683) for which
a hysteresis effect was first pointed out by Miyamoto et.
al. (1995). The difference in flux of a factor of 100 was deduced from
combining results for GX 339-4 (outburst 1988, 1991) and for 
GS 1124-683. The difference in luminosity might vary from outburst to
outburst, but data of the recent outbursts yield much lower
differences in luminosity between the spectral changes hard-soft and soft-hard.
Hard-state and soft-state observations in 1997 and 1999 were discussed
by Nowak et al. (2002, see also references therein). They report a transition
back to the hard state at a luminosity lower by about a factor of 3 
than that of the brightest hard state observation.

Zdziarski et al. (2004) show
that this system had about 15 outbursts from 1987 to 2004. For two
recent outbursts the state transitions could be observed in both
directions. In both cases the hard-soft transition
occurred at a higher flux level than the corresponding soft-hard 
transition. In the second, better observed, outburst, the luminosity 
difference for the two transitions is found to be about a factor of 5 
in the 1.5-5 keV flux (for details see Zdziarski et al. 2004).

\it{GS 1124-683, Nova Muscae}\\	
\rm{The data for GS 1124-683, mentioned above show that the soft-hard
transition occurs at a luminosity much lower than the maximum (1-37 keV)
of the X-ray flux. It is not clear when the hard-soft transition happened.
}

\it{XTE J1650-500:}\\	
\rm{
Rossi et al. (2004) presented results for the outburst of
2001/2002 and found that the state transitions occurred at
different luminosity levels, with a lower luminosity at the soft to hard
transition. The difference is about a factor 5 as shown in the
hardness-intensity diagram (Rossi et al. 2004, Fig.2).

\it {XTE J1550-564:}\\
\rm{
Kubota \& Done (2004) discuss the outburst of the microquasar XTE J1550-564 
in 1998.  The luminosity rise during the outburst was very
fast and the spectral change
from hard to soft was at a luminosity clearly higher than the one
at which the source finally changed back to the hard state, perhaps by a
factor around 10. The evolution of XTE J1550-564 
during its outburst in 2000 was reported by Rodriguez et al. (2003). 
The spectral index versus the 2-200 keV flux plotted over the outburst 
shows a hysteresis of a factor of three.

\it {Other sources:}\\
\rm{}
There are further sources where a hysteresis is suspected: 
1E 1740.7-2942 and GRS 1758-258 (Smith et al. 2002) and GRS 1915+105
(Klein-Wolt et al. 2002). 

\subsection{Definition of ``hysteresis'}

In our investigation we now use the  term ``hysteresis'' for the
feature that the luminosity at the transition from hard to soft
spectral state at the rise of one particular outburst is higher than
the luminosity at the reverse transition from soft to hard spectral
state in the decrease of the same outburst.

\section{The change to a hot coronal flow/ADAF at the inner radius of 
the thin disk}

Historically, different suggestions have been advanced to explain the 
accretion modes and the change of the spectrum.
In early work, Shapiro, Lightman \& Eardley (SLE) (1976) suggested a
hot optically thin flow which however is thermally
unstable. This is also the case for the ADAF-SLE solutions constructed
by Igumenshchev et al. (1998). ADAF-type two-temperature
solutions were first described by Ichimaru (1977) in order
to understand the two different
spectral states observed for Cyg X-1. He attributed the transition to
the energy budget of the plasma near the outer boundary of the disk,
the balance between plasma heating by viscous dissipation and
radiative loss, and thereby analytically derived a critical mass flow rate
for the spectral transition. Meyer \& Meyer-Hofmeister
(1994) proposed a model for a corona above a geometrically thin
standard accretion disk around compact objects, taking into account the
interaction of the two flows. In an apparently very different approach Honma 
(1996) considered the effect of a turbulent diffusive heat flux
outwards from a hot and mainly non-radiative advection-dominated
inner region to an outer cool accretion disk.  In spite of a very
different geometry and simplification Honma's model captures the same
physical effect as the one by Meyer \& Meyer-Hofmeister (1994). For a
discussion see Meyer et al. (2000a).
 
In connection with the application to X-ray binaries Narayan \& Yi
(1995) suggested that,
whenever the accreting gas has a choice between a thin disk and an
ADAF, the ADAF configuration is chosen 
(``strong ADAF principle''). This prescription makes it possible to derive a
relation between mass flow rate and disk truncation radius (compare
Fig. 8 in Narayan et al. 1998).

In a new systematic analysis Done \& Gierli\'nski (2004) use all data 
now available from Galactic binary systems to investigate the change of
spectra as a function of the accretion rate, and conclude that the
major hard-soft spectral transition is driven by a changing inner
radius of the accretion disk. In this picture one key feature is
missing: what determines the location of this inner radius?

The model proposed by Meyer \& Meyer-Hofmeister (1994) was originally 
worked out to understand the X-rays observed in cataclysmic variables.
But the evaporation process is even more important in disks around neutron
stars and black holes (Meyer et al. 2000a, 2000b). 
The corona is fed by matter of the thin disk which evaporates from
the cool layers underneath. This establishes a coronal mass flow 
which diminishes the mass flow rate in the thin disk. In the inner
region evaporation becomes so efficient that at low
accretion rates all matter flows via the corona and proceeds towards
the black hole as a purely coronal vertically extended accretion
flow. Very similar
to this model is the investigation of the vertical structure of
the corona by R\'o\.za\'nska \& Czerny
(2000); for a discussion of differences in the results see
Meyer-Hofmeister \& Meyer (2001).

\section{Computational results}  
\subsection{The equilibrium between cool disk and hot
corona}
In order to derive the coronal structure we take 
the standard equations of viscous hydrodynamics. For the 
results presented here we used the one-zone model approximation 
(Meyer et al. 2000a) in a modified version which takes into account
different ion and electron temperature and the effect of
Compton cooling and heating of coronal electrons by photons from the 
central area in the soft state (Liu et al. 2002, the Compton effect
now taken for hard and soft state, see below). 
The five ordinary differential equations describing the coronal flows 
above a disk are: (1) the equation of continuity, (2) the
$z$-component of momentum equation, (3) and (4) the energy equations
for ions and electrons and (5) the equation for the thermal conduction 
for a fully ionized plasma. These five equations have been written up in 
the paper of Liu et al. (2002) as Eqs. (2), (3), (4), (6) and (8).  
The boundary conditions are also taken as in Liu et al. (2002), Eq.(13)
and (14). The lower boundary condition determines the temperature 
and the relation between pressure and heat flux density at the bottom
of the corona and derives from
a standard relation between temperature and thermal flux in the very steep 
temperature profile (see Shmeleva \& Syrovatskii 1973, Liu et
al. 1995). The upper
boundary (free boundary) mirrors the requirement of no thermal heat 
input from outside and no artificial confinement at the top 
(i.e. allows for wind loss). We determine the five dependent variables: 
pressure, ion and electron temperature, heat flux density and the
vertical velocity $v_z$ as functions of height above the midplane at 
a given distance $R$. These vertical
structure solutions yield the mass evaporation rate.

Compared to Eqs. (4) and (6) of Liu et al. (2002) we introduced 
a factor of 1.5 in the term for the sidewise advection of energy, 
which now reads $\frac{3}{R}\rho_i v_R u_i$ for ions and
correspondingly for electrons. This takes into account the difference
of the specific energy between the mass flows entering and leaving the 
``one zone'' due to its radial dependence (cf. Meyer-Hofmeister 
\& Meyer 2003, Sect.2).

For equipartition field strength, synchrotron cooling in the
temperature ranges of $T=10^{8.7}K$, where radiation losses are
important in coronal models presented here, is less than
1/10 of the cooling by bremsstrahlung and is
neglegible. For higher values of magnetic to gas thermal energy density
it might however become important.

\subsection{Parameters chosen for the computations}

We take a black hole mass of $6M_\odot$. As 
claimed before (Liu et al. 2002), the results are actually
mass-independent 
as long as Compton heating by high-energy photons can be
neglected. For the viscosity parameter we take $\alpha=0.3$ (for the
influence of $\alpha$ on the evaporation efficiency see
Meyer-Hofmeister \& Meyer 2001 and Liu et al. 2002, for its use in
modeling of X-ray binary spectra Esin et al. 1997, for applications to 
accretion disk evolution Meyer-Hofmeister \& Meyer 1999).

\subsection{The effect of Compton cooling and heating}

The Compton cooling/heating rate per unit volume is the sum of Compton
cooling and heating (inverse Compton and Compton effect).
\begin{equation}\label{e:compt}
 q_{\rm Comp}={4k T_e-h\bar{\nu} \over m_e
c^2}n_e\sigma_T cu,
\end{equation}
with $k$ the Boltzmann constant, $T_e$ electron temperature, $h\bar{\nu}$
mean photon energy, $m_e$ electron mass, $c$ velocity of light, 
$n_e$ electron particle density, $\sigma_T$ Thomson cross 
section and $u$ the energy density of the photon field. 
For coronal electrons above a thin disk, the photon field is
composed of photons from the central source and those 
from the disk underneath. In our case the former are 
dominant and hence the contribution of the latter can be 
neglected here. The Compton cooling by photons of the secondary
stars is generally negligible. Even in the case of Cyg X-1 with its
supergiant companion their energy flux density at the central disk
corona is less than $10^{-3}$ of the irradiating X-ray flux density.

In the soft state where the disk reaches inward to the
last stable orbit the flux from the central region seen by the corona 
at distance $R$ is
\begin{equation}
F={L\over 4\pi R^2}{H\over R},
\end{equation}
where $L$ is the luminosity of the central
 source, which translates
into a central mass accretion rate $\dot M$ by $L=\eta \dot Mc^2$. We
use $\eta=0.1$. The factor $H/R$ is the ratio of coronal height to
distance. It takes into account that the coronal electrons only see
the central radiating disk projected on their line of sight. We take
the factor as 1/2, typical for the bulk of the coronal electrons.
For the hard spectral state we treat the inner hot ADAF region as
an external source of hard photons.
We take a lower efficiency of light production $\eta=0.05$. The
radiation comes from an inner vertically extended optically thin
region and the factor $H/R$ has to be replaced by one. 
To a corona far away from the central source all
photons come from nearly the same direction and thus $uc=F$.
For given mass accretion rate and radius the Compton cooling/heating 
rate (Eq.(\ref{e:compt})) is then a function of coronal temperature
and density. $q_{\rm Comp}>0$
means Compton cooling and $q_{\rm Comp}<0$ means heating.

For the investigation of coronal structure and evaporation rate in the
soft spectral state the mean photon energy is much less than 
the electron energy, $h\bar{\nu}\ll 4kT_e$.
For the hard spectral state we use a hard state spectrum of 100 keV
mean photon energy. Zdziarski \&
Gierli\'nski (2004) found in their analysis a range of 100 to 200
keV to be characteristic for the hard state in black hole binaries.

\subsubsection{The Compton effect in the soft and the hard spectral state}
In the soft state, Compton cooling leads to lower temperatures and
lower densities in the corona, and therefore yields a lower mass flow rate
in the corona and a lower evaporation rate than without Comptonizing 
irradiation.
Fig.1 displays three different aspects: First, the effect of Compton
cooling on the evaporation in the soft state for different 
central luminosities, i.e. mass
accretion rates ($\dot M_{\rm{Edd}}=L_{\rm{Edd}}/0.1c^2$ with 
$L_{\rm{Edd}}=4 \pi GMc/\kappa$, $\kappa$ electron scattering opacity).  
The sequence of dash-dotted, solid, dashed and dotted lines shows that
a higher central luminosity,
i.e. a higher central accretion rate leads to stronger Compton cooling
of the corona and results in less evaporation. The distance where
evaporation is maximal then moves outwards because the effect
of Compton cooling decreases with increasing distance from the
source. 

Further, Fig. 1 makes it possible to study the thin disk truncation. For 
$\dot M/\dot M_{\rm Edd}=0.01$, 
mass evaporation  (dotted lines) is very weak.  Such a mass flow rate
is higher than the maximal evaporation rate, 
$ \dot M/\dot M_{\rm{Edd}} \ge 10^{-2.86}$, so that the gas accretes to the
center through the thin optically thick disk. With 
$\dot M/\dot M_{\rm Edd}$ decreased, e.g. $\dot M/\dot M_{\rm
Edd}=0.006$, soft radiation and Compton cooling decrease,
and the evaporation rate increases
(dashed line). Also for this mass flow rate the disk would not be
truncated since evaporation cannot deplete the inner disk region.
In both these cases the disk reaches inward to the last stable
orbit with a soft multi-temperature black body spectrum.
When $\dot M/\dot M_{\rm Edd}$ decreases to 0.005, the
evaporation rate nearly reaches the same value (solid line). The
supply of accreting gas to the inner disk and hence the supply
of soft photons is stopped. The inner disk is finally depleted by
accretion and evaporation, the accretion changes from the disk-dominated to 
the RIAF/ADAF-dominated mode, and the spectrum becomes hard.

\begin{figure}
\centering
\includegraphics[width=8.8cm]{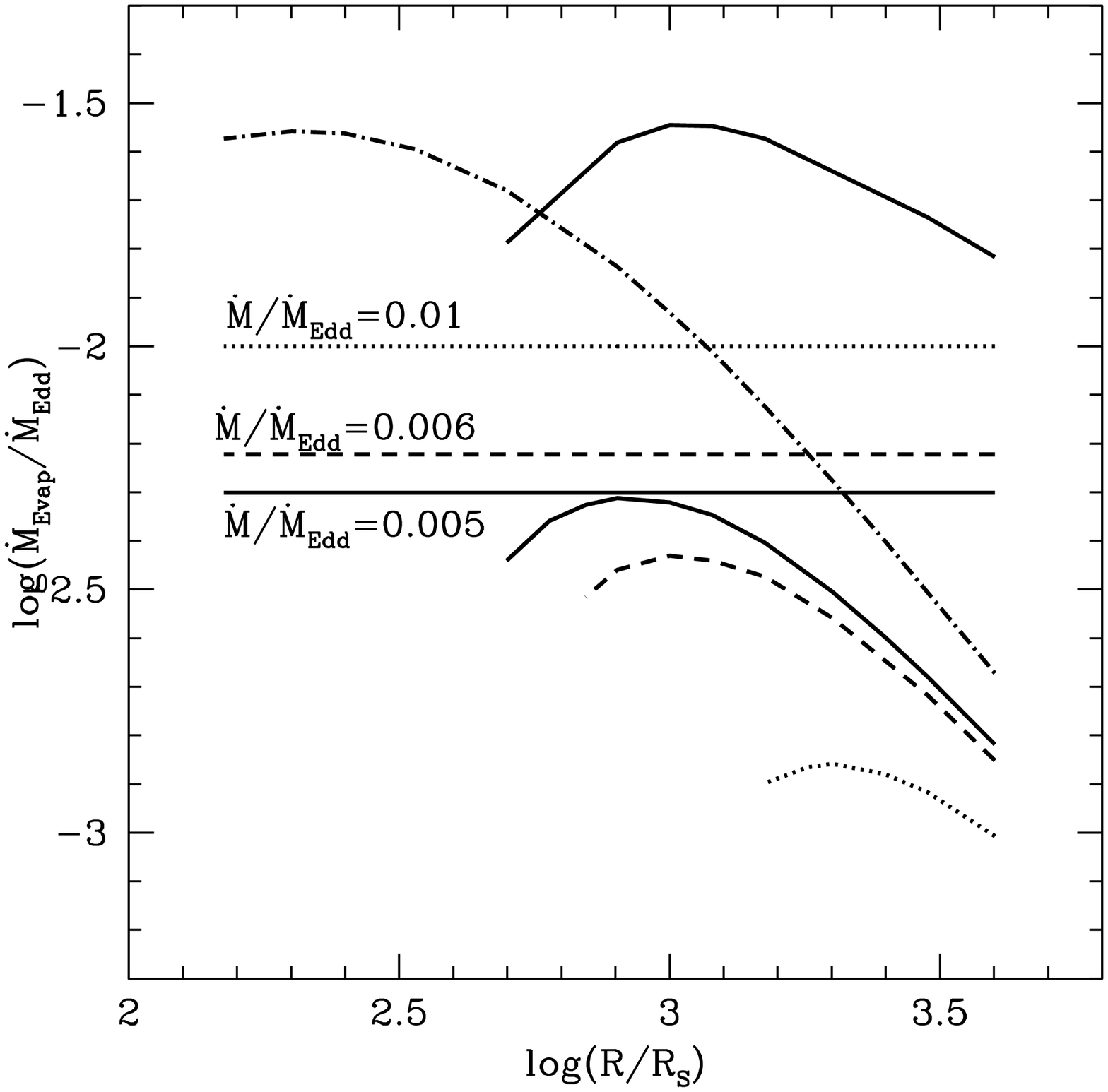}
\caption[]{Determination of the accretion rate at spectral transition
in soft and hard state.\\
(1) Soft state: Sequence of 4 curved lines: dash-dotted, solid, dashed
and dotted line: 
evaporation rates without, and with Compton effect from a soft
disk spectrum for central accretion rates $\dot M/\dot M_{\rm Edd}=
0.005,0.006, 0.01$ respectively.
Horizontal lines: Accretion rates taken for the Compton effect. 
Comparison with the evaporation rate maximum shows: for 0.01 and 0.006 
the evaporation rate is low enough to allow disk accretion
to continue inward, the spectrum is soft; for 0.005 
$\dot M/\dot M_{\rm Edd}$ this rate and the maximal evaporation rate become 
approximately equal, i.e. evaporation begins to deplete 
the disk, the transition to the hard state is triggered. \\
(2) Hard state: Solid line in the upper right diagram: evaporation rate for the
Compton effect of a hard spectrum of 100 keV mean photon energy 
for $\dot M/\dot M_{\rm Edd}=0.028$: only one curve is drawn,
evaporation for the rate of hard-soft transition.  
}
\end{figure}

Finally in Fig. 1 the result from a hard state irradiation is shown (
100 keV mean photon energy and central ADAF region luminosity for 
$\dot M/\dot M_{\rm Edd}=0.028$). This accretion rate was determined 
such that it is the same as the maximal value of the evaporation rate. At a 
slightly higher accretion rate an inner disk forms, initializing
the spectral transition to the soft state. 

What accounts for the
difference between soft and hard state irradiation? The soft state
photon energy is very low compared to that of the electrons in the corona, 
and irradiation always means cooling. In the hard state the photon
energy becomes comparable to the electron energy and as the latter
decreases with increasing distance from the central source initial
cooling turns into effective heating with the resulting high
evaporation rate.

\section{The hysteresis in the spectral transitions}
Let us discuss what the computational results mean for a full outburst 
cycle in a black hole or neutron star binary.
In quiescence the mass accretion rate is very low, evaporation
cuts off the inner disk region, inside advection-dominated accretion is
dominant. When the outburst begins the mass accretion rate $\dot M$
rises, the disk truncation radius moves inward, always to the
location where the increased accretion rate and the local 
evaporation rate are equal, following the evaporation 
rate - truncation radius relation for the hard spectral state. 
This relation however peaks at 
$\dot M \approx 2.8\times 10^{-2} \dot M_{\rm{Edd}}$ (upper right
solid line in Fig.1). When $\dot M$
surpasses this value, no disk cut-off is possible anymore, 
the disk starts to diffuse inwards to the last stable orbit, and soft 
photons from this inner disk reach the corona. From then on 
we have the coronal structure corresponding to soft state.

When the outburst declines again the accretion rate drops.  As a consequence
Compton cooling weakens and evaporation increases until at 
$\dot M/\dot M_{\rm{Edd}}\approx 5 \times 10^{-3}$ the 
mass flow in the disk and evaporation become equal,  the soft-hard
transition is triggered and accretion in the inner region takes the form
of a purely coronal flow with a hard spectrum which will remain during 
the whole quiescence.

The important feature in such an outburst cycle is that the spectral 
transitions are triggered at different accretion rates depending on
from which spectral state the transition occurs. In our example
the accretion rate at the hard-soft transition thus is a factor 
of 5 to 6 higher than the one at the inverse soft-hard transition. 
If the efficiency of hard state light
production is not much smaller than the factor $\frac{1}{2}$ assumed 
here this gives a significant hysteresis in the transition
luminosities as observed.

In Fig.2 we show the computed truncation radius at different times 
during an outburst cycle: A decrease of $R_{\rm{in}}/R_{\rm S}$ during 
rise to outburst until the hard-soft transition is reached, a constant value 
$R_{\rm {in}}=3R_{\rm S}$ as long as
the accretion rate is high (maybe very high), and a change back to a
large value during the soft-hard transition, followed by further increase.
At each moment of time the truncation radius belongs to the appropriate local 
evaporation rate. For that value the correct Compton effect 
consistent with the particular accretion rate always has to be included. 

Note that the detailed value derived for the hysteresis amplitude 
depends on the choice of the parameters used to describe the complex 
real situation by a one-zone model.

\begin{figure}
\centering
\includegraphics[width=8.8cm]{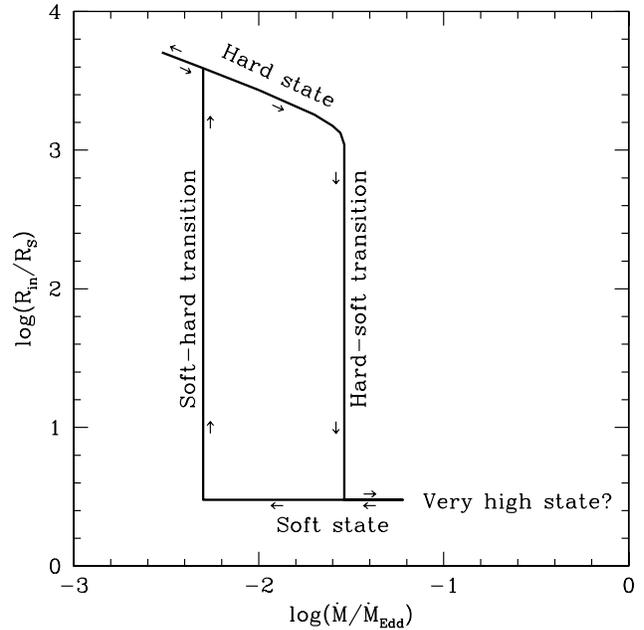}
\caption[]{The hysteresis in the truncation radius: 
The arrows show how the disk inner radius changes with mass
accretion rate during an outburst cycle (starting from the upper left
corner of the figure). In quiescence the
spectrum is hard. After onset of the outburst 
the increasing mass accretion rate yields a decreasing truncation radius.
When $\dot M/\dot M_{\rm Edd}$ has reached the 
critical value for the hard-soft state transition the disk extends inward to
the last stable orbit. During the further rise to outburst maximum and
the following decline the system stays in the soft state
($R_{\rm{in}}=3R_S$, horizontal line in the diagram), until the 
accretion rate drops to the rate at which the soft-hard
transition occurs and the disk becomes truncated again. For even smaller rates
the disk truncation moves outward. Due to the effect of Compton
cooling, the rates for the two spectral state transitions are
different, see text.
}
\end{figure}

\section{Discussion}
\subsection{Other suggestions to explain the hysteresis}

We list here suggestions by different authors as well as an
additional possibility. The first three possible scenarios were 
outlined in the investigation of Maccarone \& Coppi (2003).

(1) The state transition luminosity from an
adiabatic accretion flow to a thin disc is higher than the
transition luminosity from a thin disc to an adiabatic flow because
interactions are more efficient in the thin disc where the mean
particle separation is smaller. \hspace {0.1cm} ---  \hspace{0.1cm}
In standard theory the ADAF state would have to change to the thin
disk accretion state when the optically thin flow ceases to exist
because radiative cooling overcomes viscous heating. This boundary is
defined by the ``strong ADAF principle''. For the opposite transition
however there is no such limit: A cool disk can exist for very low
accretion rates even if it becomes optically thin. This holds as long
as corona-disk interaction is left out.

(2) During the rapid 
luminosity rise, a geometrically thin accretion flow is not stable, 
so the geometrically thick flow persists because the system is out 
of equilibrium.  \hspace {0.1cm} ---  \hspace{0.1cm} In our picture 
an increasing mass flow rate in the 
disk (in X-ray transients caused by a disk instability) shifts the 
inner edge of the thin disk inward towards the black hole or neutron 
star. These inner disk regions are radiation-pressure-dominated. 
Recently Gierli\'nski \& Done (2004) discussed the issue whether the
disks in X-ray binaries are unstable. They find that multi-temperature
black body spectra are a good fit to the observations and that no
variability in the lightcurves is found which would indicate an instability. 
Then instability should not be the cause of the hysteresis.

(3) A time lag is present
in the soft-hard transition because the disk must be evacuated or evaporated.
\hspace {0.1cm} ---  \hspace{0.1cm} In our picture the change 
from disk accretion to a hot coronal
flow will start where the evaporation rate first exceeds
the mass flow rate in the thin disk. A ring-shaped disk region
becomes evaporated and the gas in the remaining inner thin disk can disappear 
by flowing inward in the thin disk or  evaporating to a hot flow. 
At present it is not clear how fast such a left-over inner disk
region disappears completely (see also our discussion of the situation
in Cyg X-1 in the next section). The diffusive depletion time of a
disk at the distance where the transition occurs is generally short
but the situation is complex because of corona-disk interaction, and
requires more detailed investigation.

(4) Zdziarsky \& Gierlinski (2004) note that observations indicate
that in a certain range of $L/L_{\rm{Edd}}$ both a hot accretion flow or
thin disk accretion seem possible, and suggest this might be
responsible for the hysteresis in the long-time light curves of black
hole binaries.
 \hspace {0.1cm} ---  \hspace{0.1cm} This is true but by itself does
not explain how the hysteresis comes about and how big
it is.

(5) An argument which we want to add here for clarification concerns
the fact that the efficiency of light production is different for the two
modes of accretion. In our theoretical investigation we derive a
certain \it{mass flow rate} \rm{ for which the spectral state
transition occurs. If the efficiency is lower in the 
optically thin advection-dominated mode than in the optically thick 
disk accretion we expect a luminosity increase at the time of the 
hard-soft transition and a decrease at the reverse 
transition. If one attributes the newly reached higher or lower 
luminosity to the state transition this would be higher for the 
hard-soft transition and lower for the soft-hard transition, a 
difference in luminosity in the same sense as the observed
hysteresis. How large is this luminosity difference? Observation
for Cyg X-1 show only a small difference between the two states.
Czerny \& R\'o\.za\'nska (2004) derive an accretion
efficiency as a function of the distance from the black hole. In this 
evaluation the viscosity also enters. If from their investigation 
we take the values closest to our model for the
spectral state transition we find a factor of about two between the
efficiencies in disk accretion and in the hot flow. Also, the observed
luminosity from disk accretion is reduced if we see the source
at a high inclination angle. A difference of a factor of
two would not be sufficient to explain the observed hysteresis.
}

Summarizing the ideas discussed above we can conclude that the
suggestions are either not promising for explaining the hysteresis or 
at present there is not yet a quantitative result (suggestion 3) . 

\subsection{The case of Cygnus X-1}
Cyg X-1 is one of the best studied black hole X-ray binaries. Most of 
the time the system is in a hard spectral state. A soft spectral
state was observed in 1994, 1996 and 2000-2002. During the first
observed soft state the source was not observed at lower energies
(Cui et al. 1997). In the following transitions hard and soft band fluxes
were observed. From the light curve of the 1996 state transition 
obtained from the ASM and BATSE data Zdziarski \& Gierlinski (2004, Fig. 5)
conclude that there is no noticeable hysteresis in Cyg X-1 (see also
Zdziarski et al. 2002). 

The light curve of Cyg X-1 shows quite large fluctuations in the flux
on the time scale of the state transition itself. 
In Cyg X-1 with only moderate differences in the mean mass accretion rate,
the fluctuations in the flux might be
more important than in X-ray transients with rapidly changing accretion
rates. Then the effect of different Compton heating and cooling might
be washed out by backward and forward transitions.

\section{Conclusions}
 
Our investigation is aimed at understanding the mysterious hysteresis in  
the light curve of X-ray binaries and to evaluate quantitative results
for the difference in luminosity at the hard to soft and the soft to hard
spectral transitions. This hysteresis was found in the observations
of several X-ray binaries. We have shown that this is
a natural outcome of the different Compton effect that photons from 
the central light source have on the coronal electrons and thereby on 
the coronal structure in the hard and the soft state and therefore on 
the evaporation rate.

Most observations concern black hole systems. For our results we
have determined the Compton effect in the hard spectral state for a 
spectrum of 100 keV mean photon energy (Zdziarski \& Gierli\'nski
2004). For different systems peaks of the hard spectra are also found at lower
energies, e.g. at 20-30 keV for the very bright black hole 
candidate system Cyg X-3 (Szostek \& Zdziarski 2004). 
For neutron star systems whose hard spectra typically display about
equal energy contributions over the full range from 3 to 100 keV 
(Gilfanov et al. 1998) the mean photon energy is also lower, around
20 keV. In view of the clear hysteresis shown by Aql X-1 it will be
interesting to investigate those cases as well and derive a value for the
expected hysteresis.

One may note that our model not only explains the
observed hysteresis but also yields a quantitative estimate that 
agrees with the observations. At the same time this result further confirms
the picture of the evaporation model: the interaction of a cool disk
and a corona above with a maximal evaporation efficiency determining 
the spectral transition.

\begin {acknowledgements}
B.F. Liu thanks the Alexander-von-Humboldt Foundation for the award of
a research fellowship during which this investigation was performed.
\end {acknowledgements}

\end{document}